\documentclass[runningheads]{llncs}
\usepackage{graphicx}
\usepackage{MnSymbol}
\usepackage{multirow}
\usepackage{tabularx}
\usepackage{color}
\usepackage{float}  
\usepackage{subfigure}  
\usepackage{svg}

\begin{document}
\title{Causal Image Synthesis of Brain MR in 3D}
\author{Yujia Li\inst{1, 3} \and
Jiong Shi\inst{4, 5} \and
S. Kevin Zhou\inst{1, 2, 3}}
\authorrunning{Y. Li et al.}
%
\institute{Institute of Computing Technology, Chinese Academy of Sciences, Beijing, China \and
School of Biomedical Engineering, Suzhou Institute for Advanced Research, University of Science and Technology of China, Suzhou, China \and University of Chinese Academy of Sciences \and
Division of Life Sciences and Medicine, University of Science and Technology of China, Hefei, Anhui, China \and
Department of Neurology, Institute on Aging and Brain Disorders, The First Affiliated Hospital of USTC}
\maketitle              
\begin{abstract}
Clinical decision making requires counterfactual reasoning based on a factual medical image and thus necessitates causal image synthesis. To this end, we present a novel method for modeling the causality between demographic variables, clinical indices and brain MR images for Alzheimer's Diseases. Specifically, we leverage a structural causal model to depict the causality and a styled generator to synthesize the image. Furthermore, as a crucial step to reduce modeling complexity and make learning tractable, we propose the use of low-dimensional latent feature representation of a high-dimensional 3D image, together with exogenous noise, to build causal relationship between the image and non-image variables. We experiment the proposed method based on 1586 subjects and 3683 3D images and synthesize counterfactual brain MR images intervened on certain attributes, such as age, brain volume and cognitive test score. Quantitative metrics and qualitative evaluation of counterfactual images demonstrates the superiority of our generated images.

\end{abstract}
\section{Introduction}
Clinical decision making heavily relies on pattern contrasting and longitudinal comparison of medical images. Despite countless medical images are taken everyday, these factual images fail to answer the following questions, say in brain MR imaging: \textit{What would the brain image look like if the subject did not have Alzheimer's Disease? What would be the changes of lateral ventricle in MR images if taken aspirin a month ago?} This kind of inquiry involves images in situations contradicting with the fact, \textit{i.e.}, \textbf{counterfactual}~\cite{causality}, and necessitates \textbf{causal image synthesis} (CIS). 

\begin{figure}[t]
	\centering  
	\subfigcapskip=0pt 
	\subfigure[causal graph for AD]{
		\includegraphics[width=0.45\linewidth]{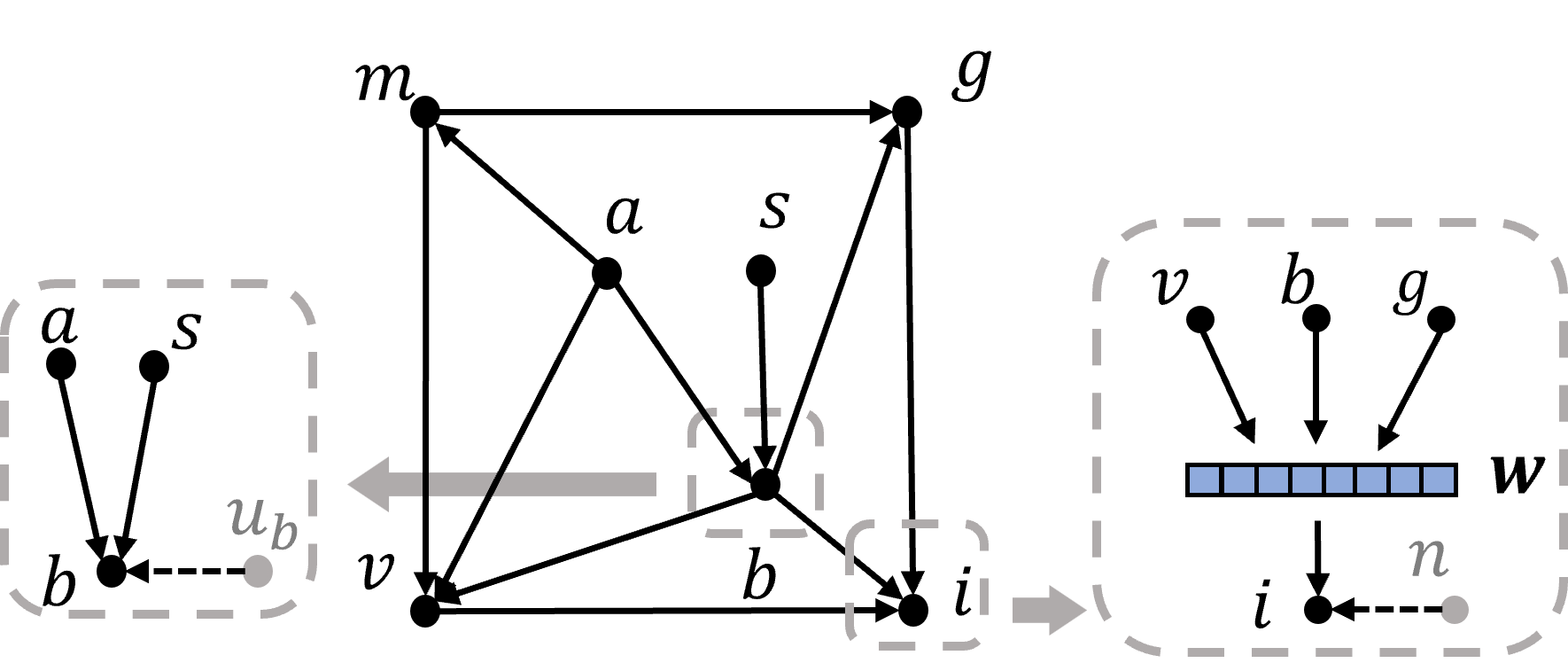}\label{fig:causalgraph}}
  \quad
	\subfigure[counterfactual MRI]{
		\includegraphics[width=0.48\linewidth]{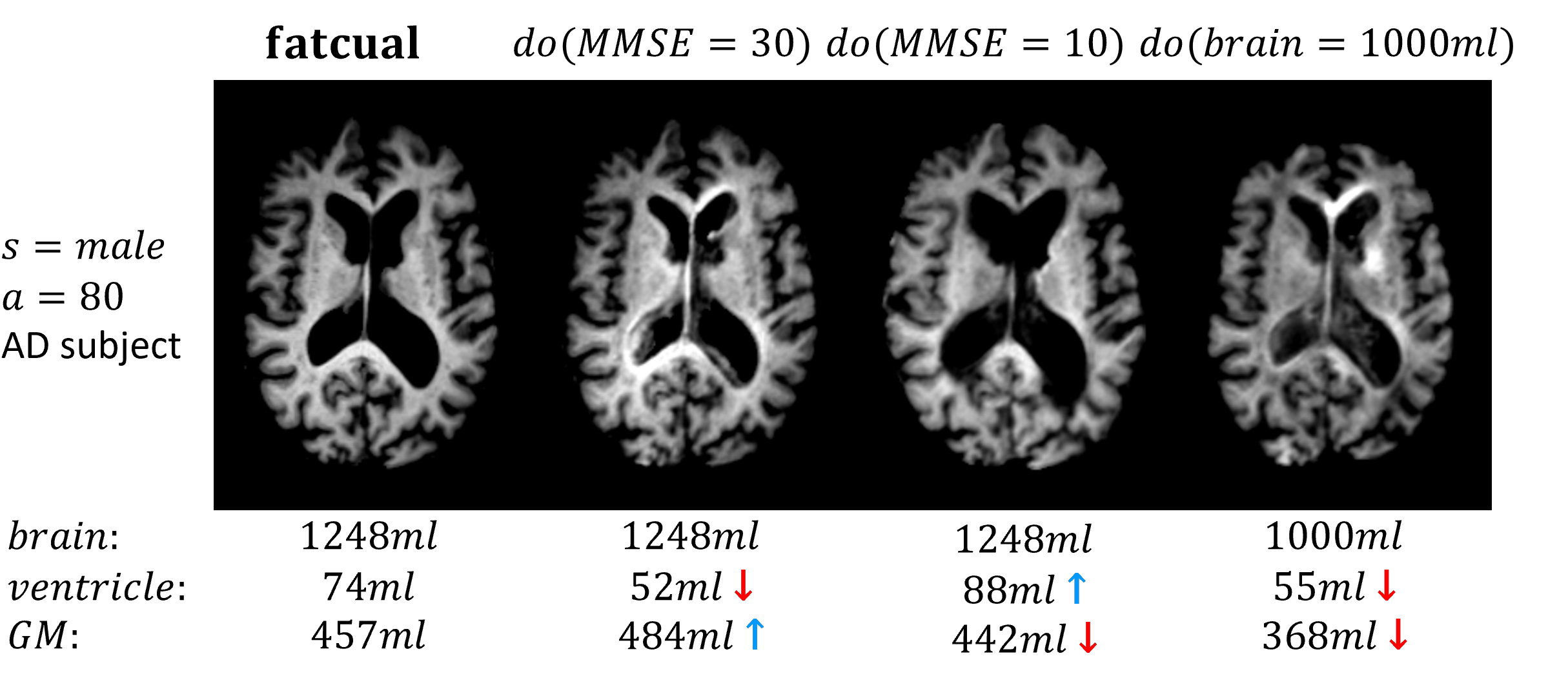}\label{fig:cf_mri}}
	\caption{(a) The causal graph for MRI for AD. Variables are image ($i$), age ($a$), sex ($s$), Mini Mental Sate Examination(MMSE) score ($m$) and brain ($b$), ventricle ($v$) , grey matter ($g$) volumes. (b) The counterfactual MRI results, set score and brain volume.}
\end{figure}

Although image synthesis has been explored extensively, existing image synthesis methods are correlation-based; yet the questions related to counterfactual images need to be answered with causality model to describe how demographic variables, disease variables and medical images interact with each other. Pearl proposes a structural causal model (SCM)~\cite{causality} which utilises a directed acyclic graph (DAG) whose nodes represent the variables and edges point from causes to effect. Take Figure \ref{fig:causalgraph} as an example, node $a$ (age) is a causal parent of node $v$ (ventricle volume), which indicates the causal effect of age on brain ventricle (increasing age causes enlarged brain ventricle). SCM has been successful in epidemiology, econometrics and medicine~\cite{epidemiology,econometrics,medicine} where the variables are in low-dimension in most cases. 

However, causal image synthesis requires handling high-dimensional image data. Furthermore, 3D medical images, \textit{e.g.}, CT, MRI, are even more challenging because of the computational demand, which leads to far less research, not to mention incorporating causality. Despite computational intractability, 3D images can provide much more information for clinical analysis. In addition, many attributes, \textit{e.g.}, volume, topological connectivity, can only be calculated in 3D. 
Thus, constructing a causal medical image synthesis model in 3D is of great importance yet challenging. In this work, we attempt to bridge such a gap. 

We choose Alzheimer's Disease (AD) as the scenario. AD is the most common type of dementia and involves the atrophy of brain, which can be reflected in MR images. Besides, the causality in AD is of plenty of research which can provide our model with prior knowledge. We manage to synthesize brain MRI in 3D while handling the causal relationships between demographic variables, clinical disease variables and MR images. This model makes it possible to (i) sample and generate controllable life-like 3D brain MR images and (ii) answer the aforementioned counterfactual questions based on existing images. Fig.~\ref{fig:cf_mri} shows the synthesised counterfactual MRI in the situation of the subject without AD or with a smaller brain, which contradicts with the fact.

Our contributions are in two folds. First, we propose a novel CIS model to generate controllable and counterfactual 3D brain MR images which can be evaluated by several metrics to ensure the reliability. To the best of our knowledge, it marks \underline{the first CIS attempt in 3D}. Second, we shed light on the relationships between demographic, disease variables and MR images for AD patients. 

\section{Related Work}
Our work is mostly related to medical image synthesis for disease modeling and other downstream tasks. A few works utilised causality for deep generative model to synthesize medical images for disease modeling~\cite{MS}, classification~\cite{classification}, segmentation~\cite{segmentation1}, but they concentrate on 2D images and lack effective metrics to evaluate the quality of synthesised images. As for 3D images synthesis, there have been works for unpaired domain translation~\cite{3D GAN me}, low-dose CT denoising~\cite{3D GAN denoise}, super-resolution~\cite{3D GAN super-resolution}, brain tumor segmentation~\cite{3D GAN segmentation} and memory-efficient GAN-based model~\cite{3D GAN HA}. However, these 3D image generative models are designed for special tasks and lacks assumptions about causal structures thus are fundamentally limited for unable to generate causal images. 

Causality in medical image analysis is also related to our work. There has been works~\cite{review1,review2} to emphasize the importance of causality for data scarcity and mismatch, fairness, domain generalization. Very limited works analyse the causality between geographical or disease variables and medical images mainly because the lack of meta-data and the insufficiency of data images themselves contain. Jiao \emph{et al}.~\cite{gene AD} discuss the causality between genetic data and Alzheimer’s Disease MRI data. Reinhold~\emph{et al.} \cite{MS} discuss the causal relationships in Multiple Sclerosis (MS) to make counterfactual predictions. However, the focus of our work is to generate causally controlled images instead of causality discovery, which allows us to apply prior medical and causality knowledge.

\section{Methods}
In this section, we first introduce SCM briefly and then the 3D StyleGAN model we use to generate MR images. Next, we explain how to combine the two to model the causality in AD and generate counterfactual MR images.

\subsection{Structural causal model}
A structural causal model (SCM)~\cite{causality} specifies a set of observed variables $\mathbf{V}$, corresponding unobserved exogenous noise $\mathbf{U}$ distributed according to $P(\mathbf{U})$ and a set of functions $\mathbf{F}$, in which $f_i$ is the causal mechanism that generates $v_i$ by $v_i = f_i(pa_i, u_i)$ with the causal parent of $v_i$ as $pa_i$ and $u_i$ the unobserved noise. A directed acyclic graph (DAG) can represent an SCM by representing variables as nodes and causal effect as edges, as illustrated in Fig.~\ref{fig:causalgraph}.

For a given SCM, $\mathbf{M} = \langle \mathbf{U}, \mathbf{V}, \mathbf{F}, P(\mathbf{U})\rangle$, under several assumptions~\cite{causality}, Pearl further provides the general solution for counterfactual inference. There are three steps: 1) $Abduction$: infer the unobserved noise $\mathbf{U}$ with observed data $\mathbf{V}$ and known functions $\mathbf{F}$; 2) $Action$: Modify the graph $\mathbf{M}$  according to desired \ intervention; and 3) $Prediction$: Compute counterfactuals $\widetilde{\mathbf{V}}$ in the new graph. 

Take the brain volume in Fig.~\ref{fig:causalgraph} as an example. The observed variable brain volume ($b$) has two causal parents: age ($a$) and sex ($s$), which along with the unobserved noise $u_b$, decide the brain volume by $v_b = f_b(pa_b, u_b), pa_b = \{v_a, v_s\}$. 
With causal mechanism $f_b$ derived and $v_b=1000 \ ml$, $v_a=50 \ years$, $v_s=0 \ (Female)$ observed, to compute the counterfactual $v_b'$ with age set to 40: 1) infer unobserved noise $p(u_b|e)$ with evidence $e = \{v_b=1000, v_a=50, v_s=0\}$;  2) set $v_a' = 40$ and compute $p(v_b') = \int p(v_b'|u_b, v_s, v_a')p(u_b|e)du$.

 
 



\subsection{Causality in AD}\label{sec.AD}
In order to generate causal MR images, firstly we need to model the causality in AD, \emph{i.e.}, to derive the causal mechanism function $f_i \in \mathbf{F}$. As shown in Fig.~\ref{fig:causalgraph}, we constructed the causal graph in AD, referring to medical research about the causality effect of age or AD on brain, ventricle and grey matter(GM) volumes~\cite{GM-AD,ventricle-AD}. Atrophy of brain and enlarged ventricle are observed for increasing age, however, the process is accelerated radically for patients with AD, especially for the areas responsible for memory
and language, mostly grey matter. We also conduct simple experiments on dataset to verify the causality, see Fig.~\ref{fig:attribute_dis} for an example. 

In Fig.~\ref{fig:causalgraph}, the geographical variables sex and age have no causal parents therefore only need to be sampled from a prior distribution. Mini Mental Sate Examination(MMSE), which is denoted by node $m$ in the graph, is a cognitive test widely used in clinical for AD. The score $m \in [0, 30]$ and a lower score indicates more cognitive impairment while health group usually score above 28. 

The $f_i$ for image $i$ is modeled by the 3D styled generator, for the $f_i$ except image, we assume that 
\begin{equation}\small
    v_i = CondAffine_{\theta}(u_i), \ u_i \sim \mathcal{N}(0, 1) \\
\end{equation}
$CondAffine_{\theta}(\cdot)$ is an affine transform, whose parameters $\theta = \boldsymbol{g}_i(pa_i)$ is a learnable, fully-connected network (FCN). Reinhold \emph{et al.}~\cite{MS} use normalizing flows (NFs) to represent $f_i$ which is more flexible but likely to overfit. 

Our assumption can be expressed in another way:
\begin{equation}\small
    v_i \sim \mathcal{N}(h_1(pa_i), h_2(pa_i)), \ h_1(\cdot)  \ and \ h_2(\cdot) \ are \ \rm FCNs.
\end{equation}
Take $v_b$ in Fig.~\ref{fig:causalgraph} as an example. For the population of female at age 60, their brain volumes satisfy a Normal distribution, whose mean and variance are determined by sex and age. For each individual, $u_b$ (noise for $v_b$) can be viewed as the individual character and further decides the precise value of $v_b$. 

Assume there are $k$ observed variables except image in the causal graph and we have observed $N$ samples, \emph{i.e.}, $N$ sets of $\{v_i\}, \ i=1,...,k$. $f_i$ are trained to maximize the probability for all variables and all samples.
\begin{equation}\small
    \{f_1, f_2, ..., f_k\} = argmax \sum_{N}\sum_{i=1}^{k} P(v_i|pa_i).
\end{equation}
With $f_i$ trained, counterfactual variables (except images) can be calculated as described in \textbf{3.1}. We put the formulas of the detailed abduction, action, and prediction for $f_i$ in supplementary material.

\begin{figure}\centering
  \begin{minipage}{0.5\linewidth}
   \subfigure[3D styled generator]{
		\includegraphics[width=1\linewidth]{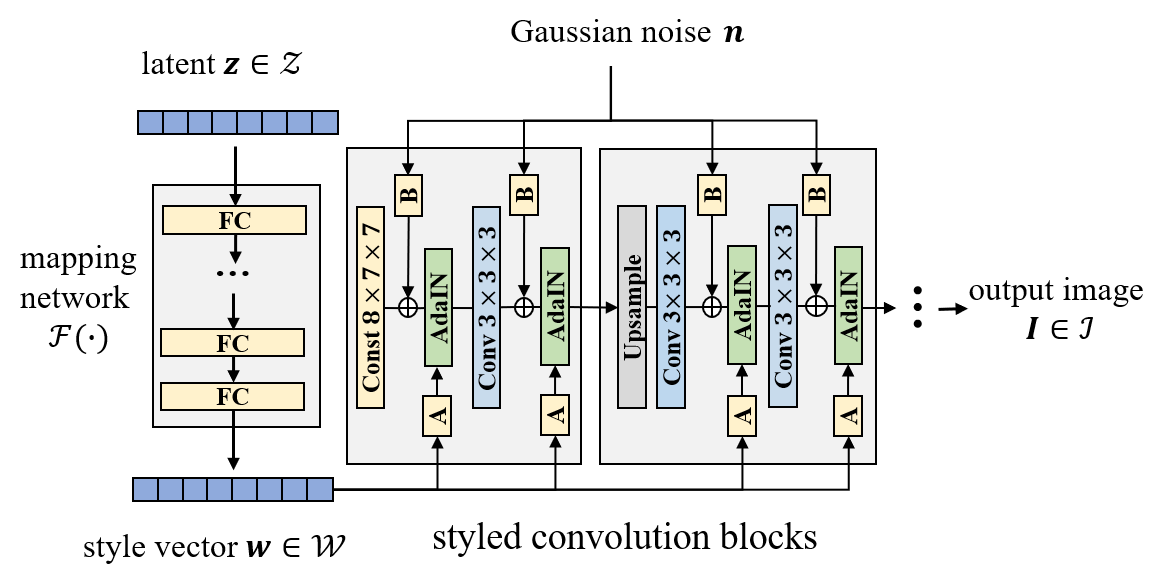}\label{fig:StyleGan}}
  \end{minipage}
  \begin{minipage}{0.45\linewidth}
   \subfigure[counterfactual MRI]{
   \includegraphics[width=1\linewidth]{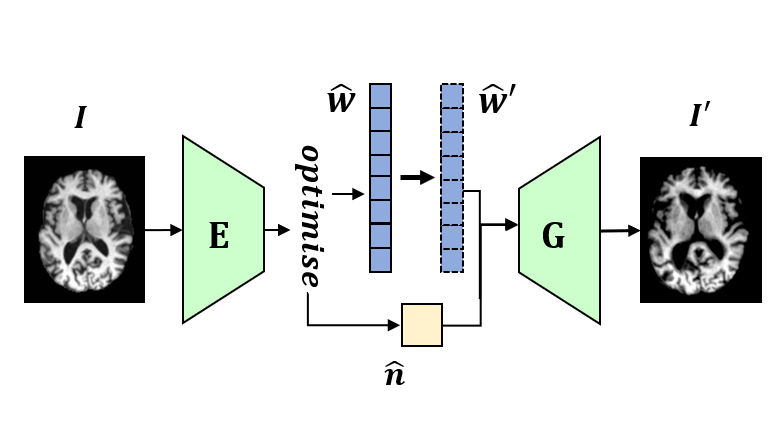}\label{fig:cf_gen_process}}
  \end{minipage}
  \caption{(a) The structure of our 3D styled generator. (b) Counterfactual MRI synthesis. Notice that $\boldsymbol{w}$ changes but the unobserved exogenous noise $\boldsymbol{n}$ remains unchanged.}

\end{figure}

\subsection{3D StyleGAN for MR image synthesis} \label{sec.style}
StyleGAN~\cite{styleGAN} is a style-based generative adversarial network that has been proved successful for its astonishing quality of generated images. StyleGAN2~\cite{styleGAN2} achieves better performance for natural images, but previous works~\cite{3D styleGAN} indicate that StyleGAN is more suitable for 3D images because of the (De)Mod operators so our 3D styled generator is on the basis of StyleGAN~\cite{styleGAN}. 

A styled generator model, as in Fig.~\ref{fig:StyleGan}, consists of a mapping network and a series of styled convolution blocks. The mapping network $\mathbf{F}$ maps latent $\boldsymbol{z} \sim \mathcal{N}(\boldsymbol{0}, \mathbf{I_{n}})$ to $\boldsymbol{w}$. The styled convolution blocks $\mathbf{G}$ then map $\boldsymbol{w}$ to image $\mathbf{I}$ with Gaussian noise $\boldsymbol{n}$ added, \emph{i.e.},
\begin{equation}
    \mathbf{I} = \mathbf{G}(\boldsymbol{w}, \boldsymbol{n}), \ where \ \boldsymbol{w} = \mathbf{F}(\boldsymbol{z}).
\end{equation}

To modify StyleGAN for 3D image,  we use 3D convolutions and change the feature map depths due to computation consideration. Like StyleGAN\cite{styleGAN}, we apply a progressive training method: train for low resolution images firstly then increase for each phase. From the beginning at $8\times7\times7$, we finally generate images of $256\times224\times224$ at a 0.8mm isotropic resolution. By contrast, S. Hong \emph{et al.}\cite{3D styleGAN} also design a 3D StyleGAN but synthesize acceptable images of $80\times96\times112$ at a 2mm resolution finally. The details of model and training can be found in the open source code~\footnote{https://anonymous.4open.science/r/CausalMRI-488C/}.

\subsection{Counterfactual image generating using styleGAN}
The causality between variables except image has been introduced in Section~\ref{sec.AD}. As shown in Fig.~\ref{fig:causalgraph}, brain ($b$), ventricle ($v$), grey matter ($g$) volumes are the causal parents of image ($i$). However, directly using the high-dimensional 3D image in SCM adds modeling complexity and makes the modeling intractable.


As introduced in Section~\ref{sec.style}, a styled generator first maps a simple Gaussian distribution space $\cal{Z}$ to $\cal{W}$ by $\mathbf{F(\cdot)}$ an then from $\cal{W}$ to image space $\cal{I}$ by $\mathbf{G(\cdot)}$. Although the training is unsupervised, many previous works have demonstrated the intriguing disentangled and semantic properties of $\cal{W}$ space both in natural image~\cite{w space1,w space2} and medical image domain~\cite{styleGAN medical2,sigmoid regression}. The disentanglement and relatively low dimension of $\cal{W}$ space provide advantages for performing image manipulation. 

Inspired by the discussion in \cite{styleGAN}, we assume that $\boldsymbol{w}$ control the basic continuous feature such as volumes in image $\mathbf{i}$ while noise $\boldsymbol{n}$ contributes to the sophisticated details, such as superfical sulcus and gyrus, as in Fig.~\ref{fig:causalgraph}. We verify the assumption by experiments in Section~\textbf{4.3}. Thus, we can preserve the character of a brain by $\boldsymbol{n}$ and bridge causality between volumes and $\boldsymbol{w}$.

To generate counterfactual MR images, as illustrated in Fig.~\ref{fig:cf_gen_process}, we need to map images $\mathbf{I}$ reverse to $\hat{\boldsymbol{w}}$ and $\hat{\boldsymbol{n}}$ with a fixed well-trained styled generator $\mathbf{G}$. We sample in the latent $\cal{Z}$ space and obtain a set of $\boldsymbol{w}$ and corresponding images $\mathbf{I}$. Then we train an encoder based on a pre-trained 3D ResNet50~\cite{Med3D} on the synthesized dataset by supervision in $\cal{W}$ space. Further optimisation is conducted for $\hat{\boldsymbol{w}}$ and $\hat{\boldsymbol{n}}$ by supervision in $\cal{I}$ space with an $L_1$ loss $|\mathbf{I}- \mathbf{G}(\hat{\boldsymbol{w}}, \hat{\boldsymbol{n}})|$. Then inspired by \cite{sigmoid regression} that utilizes a Sigmoid function for classification, we train a linear regression $y = \boldsymbol{\alpha}^{T}\hat{\boldsymbol{w}}+\beta$ for 
the volume $y$ and the encoded style vector $\hat{\boldsymbol{w}}$ to model the causality between volumes and $\boldsymbol{w}$.

Now we are ready to generate counterfactual image $\mathbf{I'} \ (set \ volume = y')$ from a factual image $\mathbf{I} \ (volume = y)$ by following three steps: 1) map $\mathbf{I}$ reverse to $\hat{\boldsymbol{w}}$ and $\hat{\boldsymbol{n}}$; 2) move $\hat{\boldsymbol{w}}$ along the direction of $\boldsymbol{\alpha}$ as volume changes; and 3) put $\hat{\boldsymbol{w}}'$ and $\hat{\boldsymbol{n}}$ into the fixed styled generator to generate the counterfactual image $\mathbf{I}'$.
\begin{equation}
    \hat{\boldsymbol{w}}' = \hat{\boldsymbol{w}} + \displaystyle{\frac{(y' - y)}{\left \|\boldsymbol{\alpha}\right \|}}\boldsymbol{\alpha}, \qquad
    \mathbf{I}' = \mathbf{G}(\hat{\boldsymbol{w}}', \ \hat{\boldsymbol{n}}).
\end{equation}
Formulas for changing multiple volumes simultaneously are in the supplement. Intervention of other variables will affect the three volumes first and then image as they are causal parents.

\section{Experiments}
\subsection{Dataset}
We used brain MR T1 images from two publicly available datasets: ADNI~\cite{ADNI1} and OASIS~\cite{OASIS2,OASIS3}. All images are skull-stripped by FSL~\cite{FSL}, registered to MNI152 space and resampled to a 0.8mm isotropic resolution by ANTs~\cite{ants}, and trimmed to 256×224×224. We then segment the processed MR images and obtain grey matter volume by ANTs and ventricle volume by a trained CNN model~\cite{ventricle seg}. Meta-data of sex, age, and MMSE score (introduced in section \textbf{3.2}) are provided by the dataset. Both healthy individuals (Health Control, HC) and cognitive impaired individuals (cognitive Impaired, CI) are included and there are multiple MR scans for every subject for at least a 6-month interval. Totally there are 1586 subjects (440 CI) and 3683 images (834 CI). 


\subsection{Causality in AD}
As discussed in Section~\ref{sec.AD}, we use $CondAffine_{\theta}(\cdot)$, $\theta = \boldsymbol{g}_i(pa_i)$, a learnable FCN to model the causality between variables except image. We train on OASIS3 (2275 images), validate on OASIS2 (373 images), and test on ADNI1 (1035 images). Compared with Normalising Flow used by \cite{DSCM,MS}, our method can avoid overfitting and better generalise as Table~\ref{tbl.logp} shows. Results on training and validation datasets are in the supplementary materials. For MMSE score, we believe that only age and sex are insufficient to generate the distribution, more such as genetic data is needed. Fig.~\ref{fig:attribute_dis} shows the true and predicted distribution of MMSE score and ventricle volume on test dataset.

\begin{table}[t]
\caption{$log(p(v_i|pa_i))$ on average. Test on OASIS2}\label{attribute}
\begin{tabular}{l p{2cm}<{\centering} p{2.5cm}<{\centering} p{1.8cm}<{\centering} p{1.5cm}<{\centering}}\\\hline
 & Brain volume & Ventricle volume & GM volume& Score  \\\hline
Normalising Flow & -1.42 & -1.50 & -1.21 & \textbf{-0.98} \\
Conditional Affine (ours) & \textbf{-1.41} & \textbf{-1.32} & \textbf{-1.10} & -2.38\\ \hline
\end{tabular} \label{tbl.logp}
\end{table}
\subsection{Synthesized 3D MRI quality}
\noindent\textbf{General synthesized image evaluation} We evaluate the synthesised images by Fréchet Inception Distance (FID) and Maximum Mean Discrepancy (MMD). Lower values of FID/MMD indicate that the distributions of synthesised images are closer to real ones. 
As shown in Table~\ref{tbl.quality}, our model is compared with the state-of-the-art 3D image generator HA-GAN~\cite{3D GAN HA} and a styleGAN model by Hong et al.~\cite{3D styleGAN}. Since they use different metrics, we take the numbers they report and compare our model with them, respectively. Figure~\ref{fig:sample mri} shows a sample and more are presented in the supplementary material.
The results prove our model can generate realistic images comparable with the-state-of-art model. We also request a clinician with over 30 years of brain image reading experience to recognize if an image is real or synthesized for 50 randomly selected images. His recognition accuracy is 50\%, exactly like flipping a fair coin.

\begin{table}\centering
\caption{Image quality evaluation. HA-GAN uses a pretrained 3D ResNet~\cite{Med3D} to extract features and compute FID and MMD, while Hong et al.\cite{3D styleGAN} compute MMD in images in miniBatch and FID of the middle slice of images (S: Sagittal, A: Axial, and C: Coronal).}\label{tab3}
\begin{tabular}{l p{1.6cm}<{\centering} p{1.6cm}<{\centering} || l p{1.3cm}<{\centering} p{0.9cm}<{\centering} p{0.94cm}<{\centering} p{0.94cm}<{\centering}}
\hline
  & FID &  MMD  & & bMMD\textsuperscript{2} & FID-S & FID-A & FID-C\\\hline
HA-GAN~\cite{3D GAN HA} & \textbf{.004(.001)} & .086(.029) & Hong et al.~\cite{3D styleGAN} & 4497(898) & 106.9 & 71.3 & 90.2 \\
ours & .010(.001)& \textbf{.024(.001)}& ours & \textbf{1109(37)} & \textbf{57.5} & \textbf{46.9} & \textbf{67.1} \\ \hline
\end{tabular} \label{tbl.quality}
\end{table}

\noindent\textbf{Counterfactual image synthesis.} The counterfactual images are evaluated for \textbf{(1)} whether they change according to the set volume, and \textbf{(2)} whether they preserve the characteristic structure of brain. We adjust the volumes by $\pm 15\%$, generate counterfactual image, calculate the volumes, and measure the SSIM between $\mathbf{I'}$ and $\mathbf{I}$. 

\begin{table} \centering
\caption{Metrics for counterfactual MRI. 
Randomly choose 500 MRI each test, generate counterfactual MRI and compute the mean and std of the changes of volumes.}
\begin{tabular}{l p{1.5cm}<{\centering} p{1.5cm}<{\centering} p{1.5cm}<{\centering} p{1.5cm}<{\centering} p{1.5cm}<{\centering} p{1.5cm}<{\centering}}
\hline
\hspace{0.4cm}Setting & -15\% &  -10\%  & -5\% & +5\% & +10\% & +15\% \\ \hline
\multicolumn{7}{c}{\textbf{Actual Change}} \\
Brain volume & -0.13(.02)&-0.11(.01)&-0.06(.01)&0.06(.01)&0.12(.01)&0.17(.02) \\ 
GM volume &  -0.13(.03)&-0.09(.01)&-0.04(.01)&0.03(.01)&0.08(.03)&0.12(.04) \\ 
Ventricle volume & -0.20(.08)&-0.13(.05)&-0.07(.01)&0.05(.01)&0.10(.03)&0.15(.07) \\ \hline
\multicolumn{7}{c}{\textbf{SSIM}} \\
Brain volume &0.78(.05)&0.85(.04)& 0.92(.02)& 0.93(.02) & 0.87(.03) & 0.81(.04)\\ 
GM volume & 0.84(.04) & 0.89(.03) & 0.94(.02) & 0.93(.02) & 0.86(.04) & 0.78(.05) \\ 
Ventricle volume &  0.94(.03) & 0.95(.02) & 0.96(.01) & 0.96(.01) & 0.96(.02) & 0.95(.03) \\\hline
\end{tabular}\label{tbl.metrics}
\end{table}

As Table~\ref{tbl.metrics} shows, the brain volume changes most accurately possibly because brain is the largest (average $1354ml$) compared with grey matter ($527ml$) and ventricle ($41ml$). SSIM between 2 random real images is 0.71(.001) and the SSIM between 2 images of the same individual (interval $>$ 1 year) is 0.85(.05), which proves that counterfactual MRI preserves the characteristic structure.  
 Figure~\ref{fig:cf_mri} visualizes counterfactual MR images. We set MMSE = 30 for an AD patient to answer the question \textit{What would the brain image look like if the subject did not have Alzheimer's Disease?} Results show a smaller ventricle and a larger grey matter volume, in accordance with the medical research on AD~\cite{GM-AD,ventricle-AD}. We also show how the image would look like if the illness is more severe (MMSE = 10) or if the subject has a smaller brain (brain volume = 1000ml). See more results in supplementary material. 
 
\begin{figure}[t]
\centering
  \begin{minipage}{0.45\linewidth}%
  \subfigure[Joint distribution of $v$ and $m$]{
		\includegraphics[width=1\linewidth]{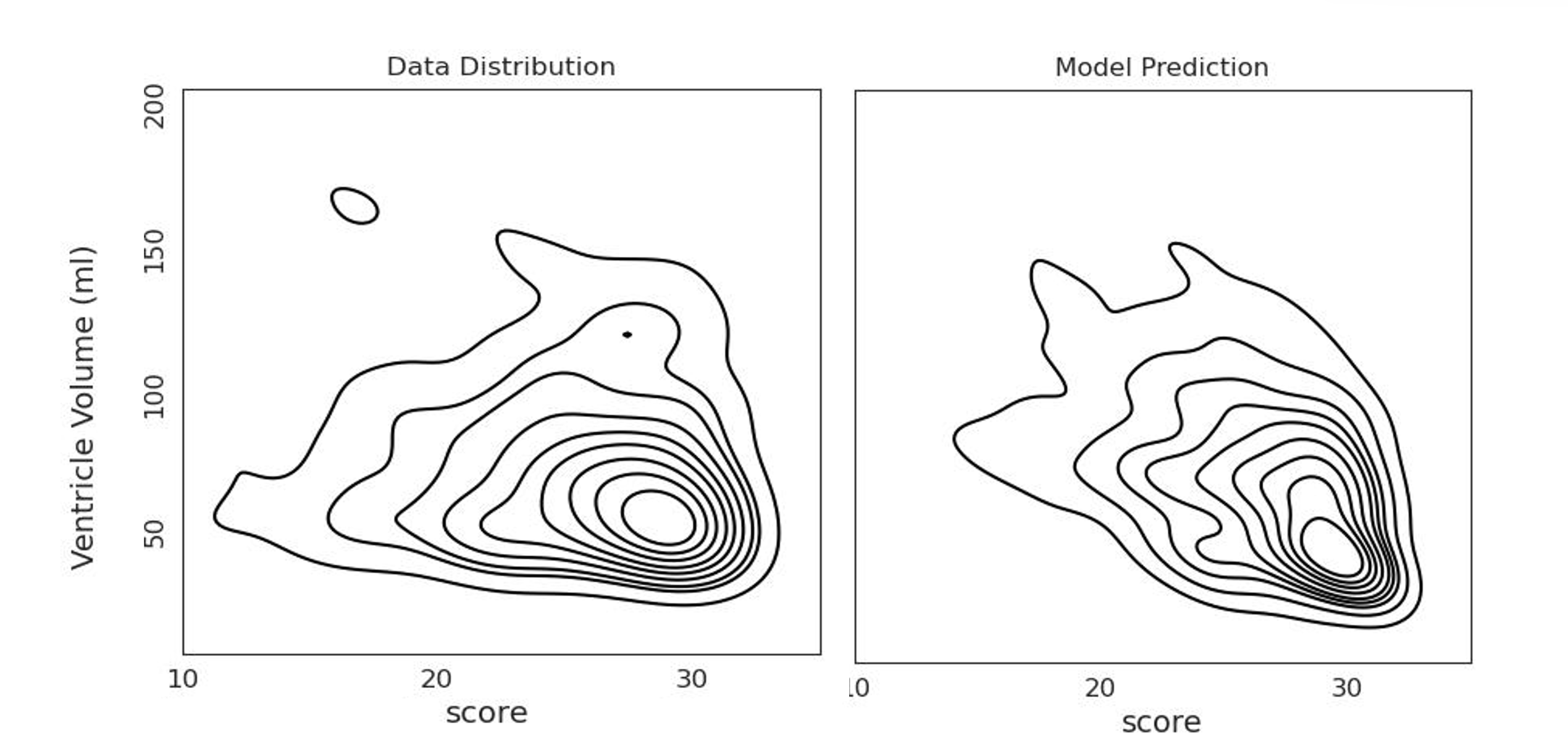}
        \label{fig:attribute_dis}}
\end{minipage}\qquad\qquad
\begin{minipage}[(b)]{0.33\linewidth}
\subfigure[Real and sampled MRI]{\includegraphics[width=1\linewidth]{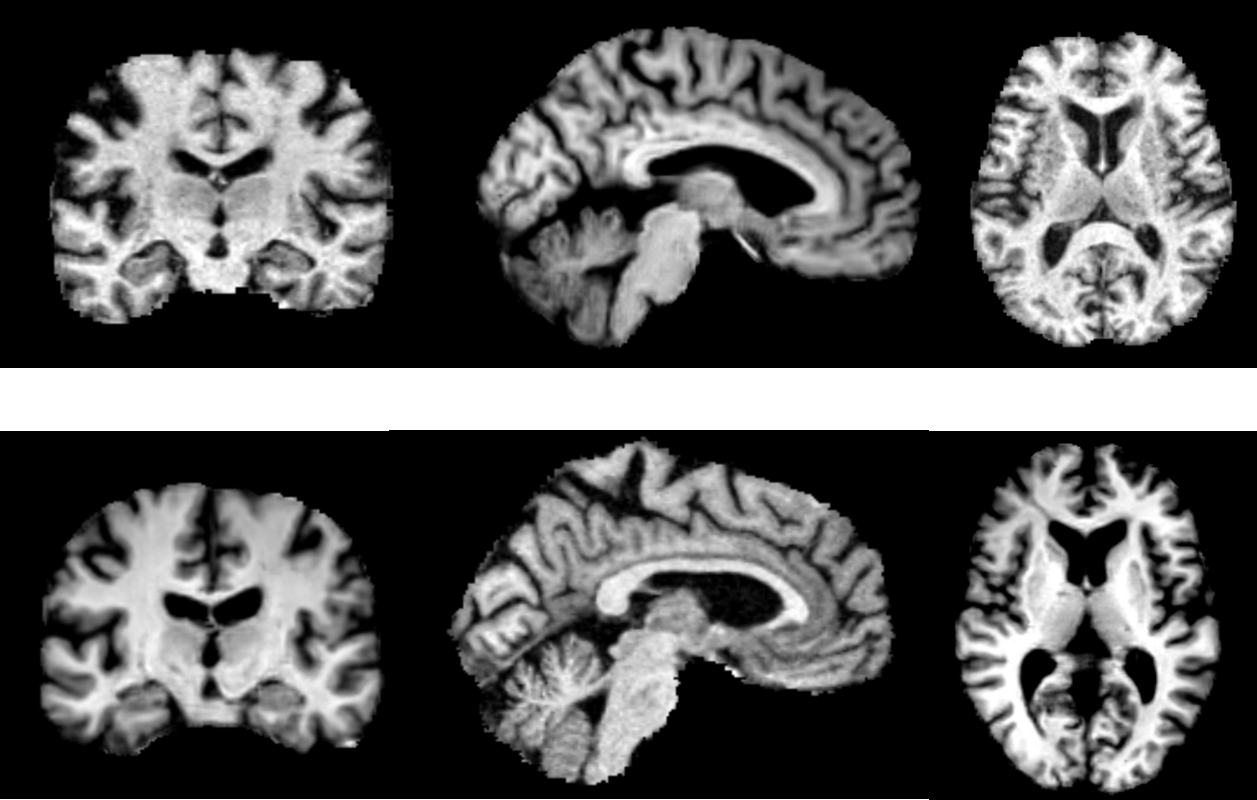}\label{fig:sample mri}}
\end{minipage}
  \caption{
  (a) The joint distribution of score and ventricle volume. Negative correlation can be observed. 
  (b) Real MRI (the first row) and synthesized MRI (the second row) sampled from
our generator.}
\end{figure}

\section{Conclusion}
This work presents a novel method for causal image synthesis in 3D. We combine structural causal model and 3D StyleGAN to depict the causality in AD and evaluate the reliability of the counterfactual images.
Our work still have some limitations. The $\mathcal{W}$ space is incapable to encode small areas such as hippocampus. This may be improved by adding supervision when training the styled generator to help $\mathcal{W}$ space learn more information. A better image generator, \textit{e.g.}, diffusion model may also address this problem.

In spite of the limitations, we believe this work has shed light on a new area: causal image synthesis. Apart from contributions to clinical analysis, a causal image synthesis model can also help to understand the relationships between images and clinical indicators. Besides, this model can augment dataset for other downstream tasks by providing controlled data as supplement, so as to benefit other image analysis model or address data bias problem.
%
%
%

\end{document}